\documentclass[%
superscriptaddress,
showpacs,preprintnumbers,
nofootinbib,
 amsmath,amssymb,
 aps,
 twocolumn,
nofootinbib
]{revtex4}
\usepackage{braket}
\usepackage{bm}
\usepackage{dcolumn}
\usepackage[dvipdfmx,final]{graphicx}
\input{colordvi.tex}
\usepackage{longtable}
\usepackage{hyperref}

\def\bea{\begin{eqnarray}}
\def\eea{\end{eqnarray}}

\def\TeV{\, {\rm TeV}}
\def\GeV{\, {\rm GeV}}
\def\MeV{\, {\rm MeV}}

\usepackage{color}

\begin{document}
\preprint{CTPU-17-13}
\title{Axion-like particle assisted strongly interacting massive particle}

\author{Ayuki Kamada}
\email{akamada@ibs.re.kr}
\affiliation{Center for Theoretical Physics of the Universe, Institute for Basic Science (IBS), Daejeon 34051, Korea}

\author{Hyungjin Kim}
\email{hjkim06@kaist.ac.kr}
\affiliation{Department of Physics, KAIST, Daejeon 34141, Korea}
\affiliation{Center for Theoretical Physics of the Universe, Institute for Basic Science (IBS), Daejeon 34051, Korea}

\author{Toyokazu Sekiguchi}
\email{sekiguti@ibs.re.kr}
\affiliation{Center for Theoretical Physics of the Universe, Institute for Basic Science (IBS), Daejeon 34051, Korea}

\date{\today}

\begin{abstract}
We propose a new realization of strongly interacting massive particles (SIMP) as self-interacting dark matter, where SIMPs couple to the Standard Model sector through an axion-like particle.
Our model gets over major obstacles accompanying the original SIMP model, such as a missing mechanism of kinetically equilibrating SIMPs with the SM plasma as well as marginal perturbativity of the chiral Lagrangian density.
Remarkably, the parameter region realizing $\sigma_{\rm self}/m_{\rm DM} \simeq 0.1 \text{--} 1 \, {\rm cm}^{2}/{\rm g}$ is within the reach of future beam dump experiments such as the Search for Hidden Particles (SHiP) experiment.
\end{abstract}


\maketitle

\section{Introduction} \label{sec:intro}

The presence of dark matter (DM) in the Universe has been firmly established by cosmological observations at scales spanning orders of magnitude, {\it i.e.}, from the cosmic microwave background (CMB) anisotropies to the rotation curve of dwarf galaxies (see, {\it e.g.}, Ref.~\cite{Bertone:2016nfn}).
However, little is known about the nature of DM as of now.
A weakly interacting massive particle (WIMP) model has been a prominent paradigm, which can be naturally accommodated in a particle physics model beyond the standard model (SM) that is suggested as a solution to the hierarchy problem ({\it e.g.}, low-scale supersymmetry~\cite{Jungman:1995df}).
However, despite all the extensive efforts so far, neither of collider experiments, direct detection experiments, nor indirect searches have found any signals of WIMPs or an underlying physics.
The WIMP paradigm is being forced into the tighter corner these days (see, {\it e.g.}, Ref.~\cite{Arcadi:2017kky}).

Besides, the conventional WIMP paradigm suffers from pressure also of observations of the matter distribution of the Universe.
The structure formation of WIMPs is concordant with that in the conventional cold DM (CDM) model and thus reproduces the observed structure at large scales.
On the other hand, there are reported discrepancies between observed subgalactic scale structures and the CDM predictions, which are collectively called the {\it small scale crisis} (see, {\it e.g.}, Ref.~\cite{DelPopolo:2016emo}).
One example is the {\it core-cusp problem} (see, {\it e.g.}, Ref.~\cite{deBlok:2009sp}): the CDM model predicts a cuspy inner density profile (inversely proportional to the distance from the center) such as the Navarro--Frenk--White profile~\cite{Navarro:1996gj} for the DM distribution in DM halos, while observed dwarf galaxies show a cored profile.
Recent state-of-the-art hydrodynamical simulations in the CDM model, which incorporate dynamical processes of baryons, rediscover the core-cusp problem in a clearer manner: a large part of mass should be expelled from an inner part of halos~\cite{Oman:2015xda}.
These observations may be hinting at DM properties that conventional WIMP models do not offer.

Self-interacting DM (SIDM) is one of the most intriguing possibilities as a solution to the core-cusp problem~\cite{Spergel:1999mh}.
With a self-scattering cross section per DM mass of $\sigma_{\rm self}/m_{\rm DM} \simeq 0.1 \text{--} 1 \, {\rm cm}^{2}/{\rm g}$, DM particles in an inner part of halos get thermalized within a dynamical time scale of halos, which leads to a lower mass density and a core DM profile~\cite{Yoshida:2000uw, Vogelsberger:2012ku, Rocha:2012jg, Vogelsberger:2015gpr}.\footnote{
The DM self-scattering also turns the shape of DM halos more spherical.
The observed ellipticity of the DM halo in galaxy clusters constrains the self-scattering cross section: $\sigma_{\rm self}/m_{\rm DM} \lesssim 0.1 \, {\rm cm}^{2}/{\rm g}$~\cite{Peter:2012jh}.
Bullet clusters imply that the colliding DM halos should pass by each other and thus put an upper bound: $\sigma_{\rm self}/m_{\rm DM} \lesssim 0.7 \, {\rm cm}^{2}/{\rm g}$~\cite{Markevitch:2003at,Randall:2007ph}.
These constraints, however, are vulnerable to modeling and/or statistical uncertainties.
In fact, $\sigma_{\rm self}/m_{\rm DM}$ inferred from a bullet cluster (Abell 3827) may be changed by orders of magnitude depending on analyses~\cite{Massey:2015dkw, Kahlhoefer:2015vua}.
Thus we take a relatively wider range of $\sigma_{\rm self}/m_{\rm DM}$.
}
SIDM also alleviates the {\it unexpected diversity problem}~\cite{Kamada:2016euw, Creasey:2016jaq} found in the aforementioned hydrodynamical simulations~\cite{Oman:2015xda}: the simulations predict similar rotation curves for similar-size dwarf galaxies, while the observed rotation curves show a diversity.
The SIDM profile is more amenable to a baryon profile that possesses a diversity even among similar-size halos, when compared to the CDM profile~\cite{Kaplinghat:2013xca, Elbert:2016dbb}.

However, particle physics aspects of SIDM are yet to be examined: how such DM can be accommodated in a concrete particle physics model as well as what a viable thermal history at an early epoch of the Universe is.
Refs.~\cite{Hochberg:2014dra, Hochberg:2014kqa} proposed a model of a strongly interacting massive particle (SIMP), where {\it pions} ($\pi$'s) in a hidden confinement sector are identified as DM.
The exact unbroken flavor symmetry may ensure the longevity of the pions.
However, there are shortcomings in the model: necessity of a kinetic equilibration mechanism and marginal perturbativity.
First, the $3 \to 2$ process induced by the Wess--Zumino--Witten (WZW) term~\cite{Wess:1971yu, Witten:1983tw} reduces the number density of pions to the observed DM mass density.
A big assumption in the Boltzmann equation of the aforementioned literature~\cite{Hochberg:2014dra, Hochberg:2014kqa} is that the DM temperature scales as that of the SM plasma.
Only with the $3 \to 2$ process and the self-scattering, however, the DM temperature scales only inversely logarithmically with that of the SM plasma~\cite{Carlson:1992fn}.
To this end, we need to maintain kinetic equilibrium between the SIMP and the SM plasma.
This is a missing piece in the original literature, which is later studied based on a {\it kinetic mixing portal}~\cite{Lee:2015gsa, Hochberg:2015vrg} and a {\it Higgs portal}~\cite{Kamada:2016ois}.
Second, the pion mass per pion decay constant is around the perturbativity bound of the na\"{\i}ve dimensional analysis ($2 \pi / \sqrt{N_{c}}$)~\cite{Manohar:1983md, Georgi:1992dw} in the parameter region where the observed DM abundance and self-interaction are obtained.
The higher-order terms of the chiral Lagrangian density may cause an order-one change in the result~\cite{Hansen:2015yaa}.

We address these issues by considering an {\it axion-like particle (ALP that we let $\phi$ denotes) portal} in the hidden confinement sector model.\footnote{
These points are also examined in different setups~\cite{Choi:2015bya, Choi:2016hid, Choi:2016tkj, Choi:2017mkk}.
}
The relic density is dominantly determined by a {\it semi-annihilation}~\cite{DEramo:2010keq} ($\pi \pi \to \pi \phi$) rather than the $3 \to 2$ process.
ALPs are well thermalized with the SM plasma and transfer kinetic energies between the pions and the SM plasma through the semi-annihilation.
We find that this mechanism works when the ALP mass is degenerate with the pion mass and the ALP decay constant is just above the electroweak scale.

This paper is organized as follows. 
In the next section, we introduce a hidden sector that incorporates an ALP as well as SIMPs, where we take into account the CP-violating terms.
In Section~\ref{sec:pheno}, we give formulas of cross sections relevant to the DM phenomenology.
Furthermore, we address how primary concerns in the original SIMP framework can be solved in our setup.
Current constraints and future detectability of our model are presented in Section~\ref{sec:const}.
We conclude in the final section.
In Appendix~\ref{sec:app}, we discuss how our model can be generalized to gauge groups other than SU$(N_{c})$.
We also provide formulas of group factors that appear in the formulas of the cross sections.

\section{Hidden sector with an ALP} \label{sec:model}
In this paper, we consider the following Lagrangian density describing the hidden sector:
\bea
{\cal L}_{\rm hid} &=& 
\label{eq:L}
\frac12 (\partial_{\mu} \phi)^{2} - V_{\rm UV}(\phi)
+ N_{L}^{\dagger} i \bar{\sigma}^{\mu} D_{\mu} N_{L}  + \bar{N}_{L}^{\dagger} i \bar{\sigma}^{\mu} D_{\mu} \bar{N}_{L} \nonumber \\
&& - m_{N} ( \bar{N}_{L} N_{L} + {\rm h.c.})
- \frac14 H^{i \mu \nu} H^{i}_{~ \mu \nu} \nonumber\\
&& + \frac{g_{H}^{2}}{32\pi^{2}} \left( \frac{\phi}{f} + \theta_{H} \right) H^{i}_{~ \mu \nu} \widetilde{H}^{i \mu \nu} ,
\eea
where $N_{L}$ and $\bar{N}_{L}$ are $N_{f}$-flavored ($N_{f} \ge 3$) vector-like fermion pairs, which respectively transform as the fundamental and anti-fundamental representations of a SU$(N_{c})$ ($N_{c} \ge 2$) gauge group, $H^{i}_{\mu \nu}$ ($\widetilde{H}^{i}_{\mu \nu}$) is the (dual) field strength of a hidden gauge field. 
The decay constant of an ALP ($\phi$) is denoted by $f$ and $V_{\rm UV}(\phi)$ is a contribution to the potential of $\phi$ from an underlying model.
It is convenient to make a chiral rotation of $N_{f}$ vector-like fermions to eliminate the theta angle in front of $H^{i}_{~ \mu \nu} \widetilde{H}^{i \mu \nu}$.
After such a chiral rotation, we see that the mass matrix becomes $m_{N} \rightarrow m_{\theta} = m_{N} e^{i \theta_{H} / N_{f}}$.
We assume that the gauge interaction confines the vector-like fermions below some energy scale ($\mu$), which is sufficiently larger than the fermion mass $m_{N}$ (we define $\theta_{H}$ so that $m_{N}$ is real).
The fermions form a condensate so that the flavor symmetry breaks as ${\rm SU}(N_{f})_{V} \times {\rm SU}(N_{f})_{A} \to {\rm SU}(N_{f})_{V}$.
Let us again remark that the unbroken flavor symmetry, SU$(N_{f})_{V}$, is essential for longevity of pions.
We parameterize the fermion bilinear as $N_{Li} {\bar N}_{Lj}  = \mu^{3} \widetilde{U}_{ij}$ with a U$(N_{f})$-valued field of 
\bea
\label{eq:piondef}
\widetilde{U} = U \exp\left[ 2 i \frac{\eta'}{f_{\eta'}} \sqrt{\frac{2}{N_{f}}} \right] , \quad U = \exp\left[ \frac{2 i \pi^{a}T^{a}}{f_{\pi}} \right] ,
\eea
where we introduce Nambu-Goldstone bosons ($\pi$'s and $\eta'$) and their decay constants ($f_{\pi}$ and $f_{\eta'}$).
Here $T^{a}$ ($a=1, \cdots, N_{f}^{2} - 1$) are the generators of SU$(N_{f})_{A}$ normalized as ${\rm Tr} (T^{a} T^{b}) = 2 \delta_{ab}$.

It is expected that the chiral anomaly provides a potential for the linear combination of $\phi / f - 2 \sqrt{2 N_{f}} \eta' / f_{\eta'}$ and the resultant $\eta'$ mass is higher than pions\,\cite{DiVecchia:1980yfw, Witten:1980sp}.
After integrating out $\eta'$, we obtain the following chiral Lagrangian density:
\bea
\label{eq:chiralL}
{\cal L}_{\rm hid} =
\frac{f_{\pi}^{2}}{16} {\rm Tr} \left( \partial_{\mu} U \partial^{\mu} U^{\dagger} \right)
+ \mu^{3} {\rm Tr} \left( m_{\theta} \widetilde{U} + {\rm h.c.} \right)
+ {\cal L}_{\rm WZW} ,
\eea
with $\widetilde{U} = U e^{i \phi/(N_{f} f)}$.
The last term is called the WZW term~\cite{Wess:1971yu, Witten:1983tw},
which introduces the $3 \leftrightarrow 2$ interaction of pions~\cite{Hochberg:2014kqa}.

There is no reason why the minimum of the axion UV potential is aligned with that of the potential originating from $H^{i}_{~ \mu \nu} \widetilde{H}^{i \mu \nu}$ and the CP symmetry is respected. 
Once we assume that $\phi$ dominantly obtains the mass from $V_{\rm UV}(\phi)$, the theory violates CP symmetry. 
We define $\theta_{H}$ so that $V_{\rm UV}(\phi)$ takes the minimum at $\phi = 0$, and then the order parameter for CP violation is given as Im$ (m_{\theta}) \propto \sin \left( \theta_{H} / N_{f} \right) $.
We remark that periodicity of $\theta_{H} \to \theta_{H} + 2 \pi n$ ($n$: integer) is maintained in the chiral Lagrangian since $\exp (2 \pi i n / N_{f})\in$ SU$(N_{f})$ and thus can be eliminated by a chiral transformation of $\pi$'s.

We expand the matrix of $\widetilde{U}$ to obtain the following Lagrangian density of the pions and the ALP,
\bea
\label{eq:perchiralL}
{\cal L}_{\rm hid} = {\cal L}_{0} + {\cal L}_{\rm CP} +{\cal L}_{\rm CPV} + {\cal L}_{\rm WZW} ,
\eea
where 
\bea
\label{eq:WZW}
{\cal L}_{0} &=&
\frac12 \left( \partial_{\mu} \pi^{a} \right)^{2} + \frac12 \left( \partial_{\mu} \phi \right)^{2} - \frac12 m_{\pi}^{2} \left( \pi^{a} \right)^{2} - \frac12 m_{\phi}^{2} \phi^{2} , \nonumber \\
{\cal L}_{\rm CP} &=& \frac{m_{\pi}^{2}}{4 N_{f}^{2} f^{2}} \left( \pi^{a} \right)^{2} \phi^{2} - \frac{1}{6 f_{\pi}^{2}} r_{abcd} \left( \partial_{\mu} \pi^{a} \right) \left( \partial^{\mu} \pi^{b} \right) \pi^{c} \pi^{d}  \nonumber\\
&& + \frac{m_{\pi}^{2}}{6 N_{f} f_{\pi} f} d_{abc} \pi^{a} \pi^{b} \pi^{c} \phi + \frac{m_{\pi}^{2}}{12 f_{\pi}^{2}} c_{abcd} \pi^{a} \pi^{b} \pi^{c} \pi^{d} , \nonumber\\
{\cal L}_{\rm CPV} &=& \tan \left( \theta_{H} / N_{f} \right) \left[ \frac{m_{\pi}^{2}}{2 N_{f} f} \phi (\pi^{a})^{2} \right. \nonumber \\
&& \left. + \frac{m_{\pi}^{2}}{6 f_{\pi}} d_{abc} \pi^{a} \pi^{b} \pi^{c} - \frac{m_{\pi}^{2}}{30 f_{\pi}^{3}} \pi^{a} \pi^{b} \pi^{c} \pi^{d} \pi^{e} c_{abcde} \right] , \nonumber \\
{\cal L}_{\rm WZW} &=& \frac{2 N_{c}}{15 \pi^{2} f_{\pi}^{5}} \epsilon^{\mu \nu \rho \sigma} c_{[abcde]} \left( \pi^{a} \partial_{\mu} \pi^{b} \partial_{\nu} \pi^{c} \partial_{\rho} \pi^{d} \partial_{\sigma} \pi^{e} \right) .
\eea
Here, we define $m_{\pi}^{2} f_{\pi}^{2} = 16 m_{N} \mu^{3} \cos (\theta_{H} / N_{f})$. 
The axion mass ($m_{\phi}$) also receives a contribution from the UV potential as mentioned above: $m_{\phi}^{2} \geq m_{\pi}^{2} f_{\pi}^{2} / (8 N_{f} f^{2})$.
In the above expression, we have kept only relevant terms.
Group factors such as $d^{2}$ are defined as given in Eqs.~\eqref{eq:groupfac2}--\eqref{eq:groupfac} in Appendix~\ref{sec:app}.
We let a (square) parenthesis in a sub/superscript indicate the total (anti-)symmetrization of the enclosed indices.

In addition, we assume that $\phi$ couples to the SM sector via the following Lagrangian density:
\bea
\label{eq:FFtilde}
{\cal L}_{\phi \gamma \gamma} = C_{\phi \gamma \gamma} \frac{\alpha}{4 \pi} \frac{\phi}{f} F_{\mu\nu} {\widetilde F}^{\mu\nu} ,
\eea
where $\alpha$ is the fine structure constant and $F^{\mu\nu}$ (${\widetilde F}^{\mu\nu}$) is the (dual) field strength of the photon.
$C_{\phi \gamma \gamma}$ is a constant typically of order unity depending on an underlying model.

\section{Dark matter phenomenology} \label{sec:pheno}
First, let us compute the flavor-averaged cross sections of the following processes relevant to the pion freeze-out in the early Universe and their self-interactions in DM halos at a later era: the semi-annihilation ($\pi \pi \to \pi \phi$), the $3 \to 2$ process ($\pi \pi \pi \to \pi \pi$), and the self-scattering ($\pi \pi \to \pi \pi$).
We assume that the initial (left-hand side) pions are non-relativistic. 
Next, we describe the roles the semi-annihilation plays during the pion freeze-out, especially stressing that the semi-annihilation contributes to the kinetic equilibration between the pions and the SM plasma when the masses of the pions and the ALP are degenerate: $m_{\pi} \simeq m_{\phi}$.
We also show to what extent the semi-annihilation helps us to mitigate the perturbativity issue.

The cross section of the semi-annihilation ($\pi \pi \to \pi \phi$) is given by
\bea
\label{eq:semisigma}
\langle \sigma_{\rm semi} v_{\rm rel} \rangle &\simeq& \frac{1}{64\pi} \frac{m_{\pi}^{2}}{f_{\pi}^{2} f^{2}} \frac{d^{2}}{N_{f}^{2} N_{\pi}^{2}} {\cal I}(m_{\phi},m_{\pi},T) \nonumber \\
&& \times \left[ 1 + \frac{m_{\pi}^{2} + m_{\phi}^{2}/9}{m_{\pi}^{2} - m_{\phi}^{2}/3} \tan^{2} \left( \theta_{H} / N_{f} \right) \right]^{2} ,
\eea
where $d^{2} = \sum_{abc} d_{abc}^{2}$ [see also Eqs.~\eqref{eq:groupfac2}--\eqref{eq:groupfac} in Appendix~\ref{sec:app}].
The brackets denote the thermal average with temperature of $T$.
If $m_{\phi} \ll m_{\pi}$, the phase space factor ${\cal I}(m_{\phi},m_{\pi},T)$ is given by
\bea
\label{eq:Imasslesslim}
{\cal I} =
\frac34
\sqrt{\left( 1- \frac{m_{\phi}^{2}}{9 m_{\pi}^{2}} \right) \left( 1 - \frac{m_{\phi}^{2}}{m_{\pi}^{2}} \right) }
\left[ \frac{K_{1}(m_{\pi} / T)}{K_{2} (m_{\pi} / T)} \right]^{2} ,
\eea
where $K_{n}(x)$ is the $n$th-order modified Bessel function of the second kind.
On the other hand, if the masses are degenerate ($m_{\phi} = m_{\pi}$), we find
\bea
\label{eq:Idegegeratelim}
{\cal I} = \frac{2 T}{m_{\pi}} \frac{K_{1}(2 m_{\pi} / T)}{K_{2}^{2}(m_{\pi} / T)} ,
\eea
which can be approximated as ${\cal I} \approx 2/\sqrt{\pi x} = v_{\rm rel}/2$ for $x = m_{\pi} / T \gg 1$.

The cross sections of the $3 \to 2$ process ($\pi \pi \pi \to \pi \pi$) and the self-scattering ($\pi \pi \to \pi \pi$) are found in the original model~\cite{Hochberg:2014kqa}, but they need to be extended to incorporate a nonzero CP phase. 
We obtain
\bea
\label{eq:cannsigma}
\langle \sigma_{3 \to 2} v_{\rm rel}^{2} \rangle &=& \frac{5 \sqrt{5}}{2\pi^{5}} \frac{N_{c}^{2} m_{\pi}^{5} t^{2}}{f_{\pi}^{10}N_{\pi}^{3} x^{2}} \nonumber \\
&& + \frac{\sqrt{5}}{2304 \pi} \tan^{2} \left( \theta_{H} / N_{f} \right) \frac{m_{\pi}}{N_{\pi}^{3} f_{\pi}^{6}} \{CD\}^{2} , \\
\label{eq:selfsigma}
\sigma_{\rm self}
&=& \frac{1}{32\pi} \frac{m_{\pi}^{2}}{N_{\pi}^{2} f_{\pi}^{4}} \left[ \{C+R\}^{2} - \tan^{2} \left( \theta_{H} / N_{f} \right) \right. \nonumber \\ 
&& \times \{C+R\}D^{2} +  \tan^{4} \left( \theta_{H} / N_{f} \right) \frac{D^{4}}{4} \left. \right] ,
\eea
where again group factors such as $t^{2}$ are defined in Eqs.~\eqref{eq:groupfac2}--\eqref{eq:groupfac} of Appendix~\ref{sec:app}.

Let us recall that $f$ should be substantially larger than $f_{\pi}$ so that the Lagrangian density in Eq.~\eqref{eq:L} is valid.
It follows that the other processes such as the annihilation of a pion pair into the ALPs ($\pi \pi \to \phi \phi$) and the scattering of the pion with the ALP ($\pi \phi \to \pi \phi$) are not relevant in the course of the pion freeze-out.
For example, $\pi \pi \to \phi \phi$ contributes to the chemical equilibration between the pions and the ALPs, but decouples earlier than $\pi \pi \to \pi \phi$, since the cross section is suppressed by $(f_{\pi}/f)^{2}$ when compared to $\langle \sigma_{\rm semi} v_{\rm rel} \rangle$. 
For the same reason, we find that $\pi \phi \to \pi \phi$ is not efficient enough to keep the pions in kinetic equilibrium with the SM plasma during the pion freeze-out.

Whether the semi-annihilation or the $3 \to 2$ process dominates the chemical equilibration depends on the masses and the decay constants of the pions and the ALP.
Either process, in general, leads to a conversion of the DM mass energy to the kinetic one. 
Unless DM particles can efficiently deposit the injected kinetic energy into the SM plasma, DM particles are heated up in the course of the pion freeze-out. 
In such a case, kinetic equilibrium is hardly maintained between the DM particles and the SM plasma, so that the evolution of the DM temperature becomes far nontrivial.
Although the temperature evolution out of kinetic equilibrium is worth investigating, we leave this for a future study~\cite{prep}, and in the rest of this paper we focus on the case that the masses are degenerate between the pions and the ALP, {\it i.e.}, $m_{\pi} \simeq m_{\phi}$ and the semi-annihilation dominates the $3 \to 2$ process.
Assuming the degenerate masses, we omit the conversion of a mass deficit to the kinetic energy from the semi-annihilation. 
As a consequence, the semi-annihilation now contributes to the kinetic equilibration between the DM particles and the SM plasma as well as the chemical one.
Thus, the DM freeze-out in our scenario proceeds in the same manner as the semi-annihilating DM model (see, {\it e.g.}, Ref.~\cite{DEramo:2010keq}).
Furthermore, as we will see closely in the next section, the degenerate masses help our DM pions to evade constraints from indirect searches for the semi-annihilation at a later epoch of the Universe. 

The domination of the semi-annihilation also helps us to alleviate the issue regarding perturbativity in the original model.
To see this, we take $N_{f}=4$, $\theta_{H}=0$, $\alpha_{\pi} = m_{\pi}/f_{\pi} = 2$, and $f = 200 \GeV$ as a benchmark point (denoted by $\star$ in Fig.~\ref{fig:constraints}).
At this benchmark point, we obtain
\bea
\langle \sigma_{\rm semi} v_{\rm rel} \rangle|_{T =T_{\rm fo}} &\simeq& \frac{6\times 10^{-9}}{{\rm GeV}^{2}} \left( \frac{\alpha_{\pi}}{2} \right)^{2} \left( \frac{200 \GeV}{f} \right)^{2} \nonumber \\
&& \times \sqrt{ \frac{19}{x_{\rm fo}} } , \nonumber \\
\langle \sigma_{3\rightarrow2} v_{\rm rel}^{2} \rangle|_{T =T_{\rm fo}} n_{\rm fo} &\simeq& \frac{4\times 10^{-11}}{\GeV^{2}} \left( \frac{\alpha_{\pi}}{2} \right)^{10} \left( \frac{100 \MeV}{m_{\pi}} \right)^{6} \nonumber \\ 
&& \times \left( \frac{N_c}{3} \right)^{2} \left( \frac{19}{x_{\rm fo}} \right)^{2} , \nonumber \\
\frac{\sigma_{\rm self}}{m_{\pi}} &\simeq& 0.9 \, {\rm cm}^{2} / {\rm g} \left( \frac{\alpha_{\pi}}{2} \right)^{4} \left( \frac{100 \MeV}{m_{\pi}} \right)^{3} , \nonumber
\eea
where $n_{\rm fo}$ and $T_{\rm fo}$ respectively denote the freeze-out number density, which is determined by the observed DM density, and temperature at the pion freeze-out. 
For $\alpha_{\pi} = 2$ (below the perturbativity bound) and $\sigma_{\rm self} / m_{\pi} = 1 \, {\rm cm}^{2}/{\rm g}$ (the SIDM cross section), the cross section of the $3 \to 2$ process ($\pi \pi \pi \rightarrow \pi \pi$) is too small to provide the observed DM mass density (recall the canonical WIMP cross section: $\langle \sigma v_{\rm rel} \rangle_{\rm can} \simeq 3 \times 10^{-9} / {\rm GeV}^{2}$).
The semi-annihilation cross section, on the other hand, takes the appropriate value to result in the observed DM mass density, provided that the axion decay constant is around the electroweak scale. 
Fig.~\ref{fig:constraints} shows the parameter regions where we obtain the observed DM abundance ($\Omega_{\rm DM} h^{2} \simeq 0.12$, black line), the self-scattering cross section for SIDM ($0.1 \, {\rm cm}^{2}/{\rm g} \leq \sigma_{\rm self}/m_{\pi} \leq 1 \, {\rm cm}^{2}/{\rm g}$, red hatched), and the semi-annihilation cross section going below that of the $3\to2$ process (magenta shaded). 
As long as the semi-annihilation dominates the $3 \to 2$ process, the relic abundance can be realized with the appropriate value of $f$. 
The figure apparently shows that our model reconciles perturbativity with the SIDM cross section.

Our argument on the pion freeze-out so far relies on a few implicit assumptions, which we clarify now.
First, the ALPs are assumed to be thermalized with the SM plasma at least until the freeze-out of DM.
The Primakoff process and the decay and inverse decay through the interaction given in Eq.~\eqref{eq:FFtilde} are responsible for thermalization of the ALPs. 
In particular, the decay and inverse decay are efficient in the course of the pion freeze-out.
When the ALPs are relativistic, the rate of the decay and inverse decay is approximately given by
\bea
\langle \Gamma_{\rm dec} \rangle \simeq C_{\phi \gamma \gamma}^{2} \frac{\alpha^{2}}{768 \pi \zeta(3)} \frac{m_{\phi}^{4}}{f^{2} T} ,
\eea
where $\zeta(x)$ is the Riemann zeta function.
The re-coupling temperature, below which the decay and inverse decay are efficient, is estimated at
\bea
T_{\rm rec} \simeq 2 \GeV \, C_{\phi \gamma \gamma}^{2/3} \left( \frac{m_{\phi}}{100 \MeV} \right)^{4/3} \left( \frac{200 \GeV}{f} \right)^{2/3} .
\eea
We find that thermalization of the ALPs is guaranteed in the parameter region allowed by the existing constraints that are discussed in the next section and shown in Fig.~\ref{fig:constraints}.

Second, without a further extension of the hidden sector, the pions need to be produced efficiently from the SM plasma after the Universe is reheated.
Even if the reheating temperature is very low, for example, $T_{\rm rh} \lesssim \left( 2 \pi /\sqrt{N_{c}} \right) f_{\pi}$, pions can be produced through $\phi \phi \to \pi \pi$ first.
Its cross section is given by
\bea
\langle \sigma_{\phi \phi \to \pi \pi} v_{\rm rel} \rangle &\simeq& \frac{1}{128 \pi} \frac{m_{\pi}^{2}}{N_{f}^{4} f^{4}} \left[ 1 + \tan^{2} \left( \theta_{H} / N_{f} \right) \right]^{2} \nonumber \\ 
&& \times {\cal I}(m_{\phi},m_{\pi},T) ,
\eea
when the ALPs are non-relativistic and $m_{\pi} \simeq m_{\phi}$ [see eq.~\eqref{eq:Idegegeratelim} for ${\cal I}$].
The number of the $\phi \phi \to \pi \pi$ reactions per Hubble time can be smaller than unity:
\bea
\langle \sigma_{\phi \phi \to \pi \pi} v_{\rm rel} \rangle n_{\phi} / H &\simeq& 0.03 \left( \frac{m_{\pi}}{100 \MeV} \right)^{3} \left( \frac{200 \GeV}{f} \right)^{4} \nonumber \\
&& \times \left( \frac{T}{m_{\pi}} \right)^{3/2} e^{- m_{\pi}/T} ,
\eea
where we take $N_{f}=4$ and $\theta_{H} = 0$ again. 
However remark that the efficient semi-annihilation multiplies the produced number of pions by the exponential of
\bea
\langle \sigma_{\rm semi} v_{\rm rel} \rangle N_{\pi} n_{\phi} / H &\simeq& 2 \times 10^{8} \left( \frac{\alpha_{\pi}}{2} \right)^{2} \left( \frac{m_{\pi}}{100 \MeV} \right) \nonumber \\
&& \times \left( \frac{200 \GeV}{f} \right)^{2} \left( \frac{T}{m_{\pi}} \right)^{3/2} e^{- m_{\pi}/T} . \nonumber \\
\eea
The above two observations indicate that in the viable parameter region shown in Fig.~\ref{fig:constraints}, the pions can be efficiently sourced as long as the reheating temperature is larger than $T_{\rm fo}$.

\section{Constraints and future detectability} \label{sec:const}

\begin{figure*}[!t]
\centering
\begin{tabular}{ccc}
\includegraphics[scale=0.45]{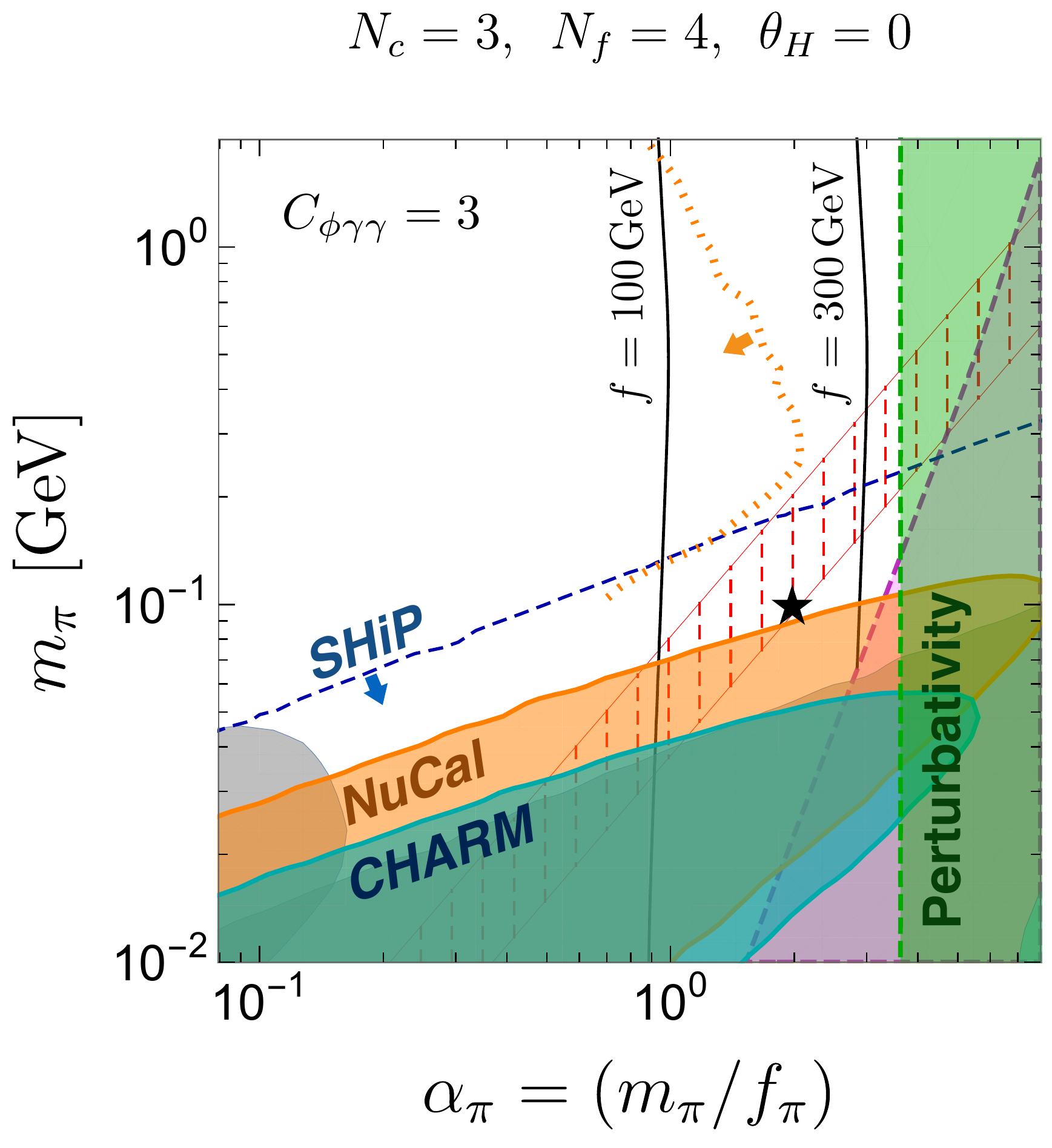} &
\qquad\qquad
\includegraphics[scale=0.45]{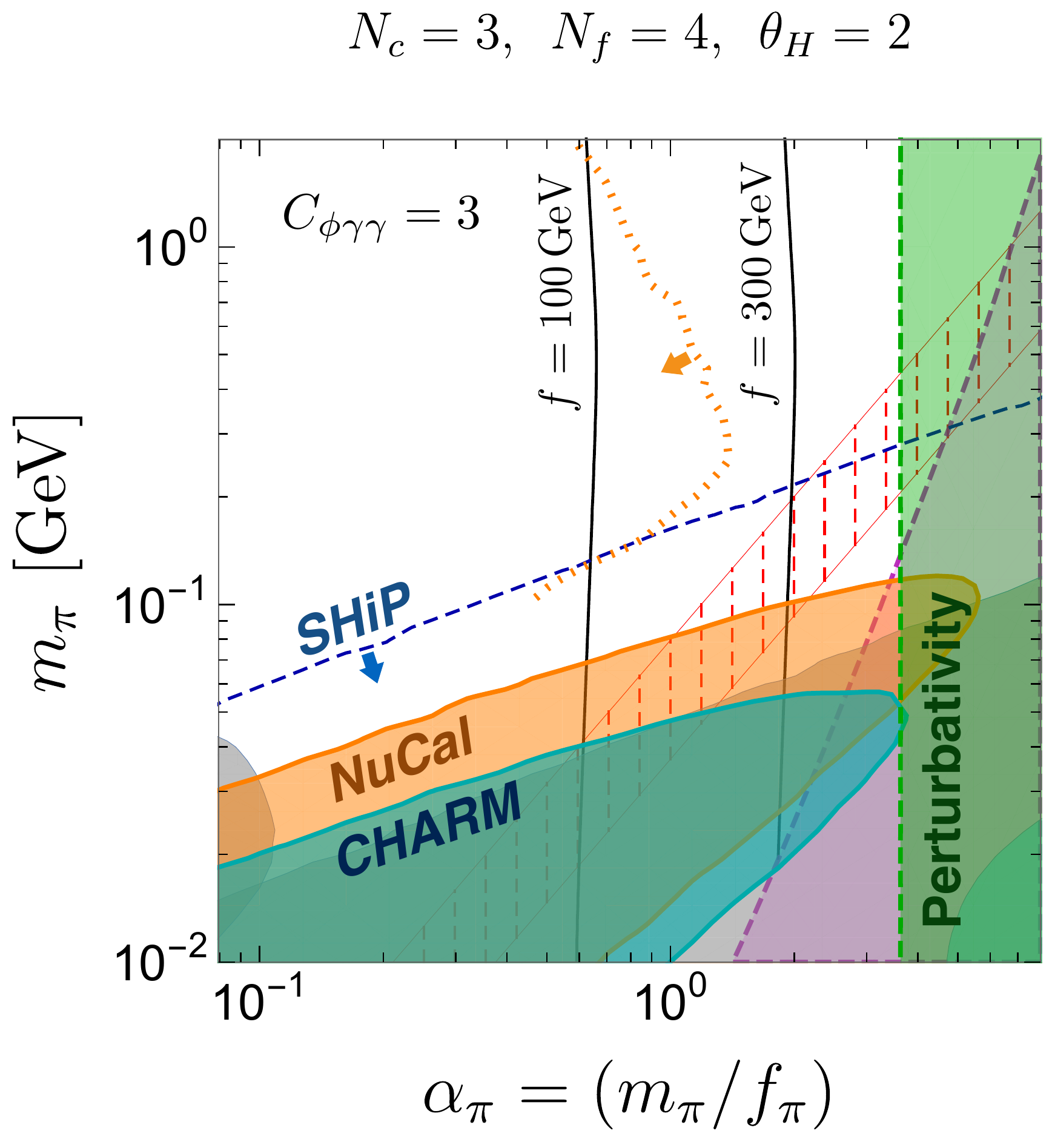}
\end{tabular}
\caption{
Shown are $m_{\pi}$--$\alpha_{\pi}$ planes of the hidden sector.
We assume $N_{c}=3$, $N_{f}=4$, and $C_{\phi \gamma \gamma}=3$ in addition to the degenerate masses: $m_{\phi}=m_{\pi}$. 
$\theta_{H}$ is taken to be $0$ (left panel) and $2$ (right panel).
Black lines give the observed DM abundance ($\Omega_{\rm DM} h^{2}=0.12$) for $f= 100$ and $300 \GeV$ from the left to the right in each panel. 
In the red hatched band, the pion self-scattering achieves the SIDM cross section, {\it i.e.}, $0.1 \, {\rm cm}^{2}/{\rm g} \leq \sigma_{\rm self}/m_{\pi} \leq 1 \, {\rm cm}^{2}/{\rm g}$. 
In the magenta shaded region, the $3 \to 2$ process dominates the semi-annihilation and thus determines the pion freeze-out.
In the green region, the pion mass exceeds a na\"{\i}ve cutoff scale of the chiral Lagrangian [see Eq.~\eqref{eq:chiralL}]~\cite{Manohar:1983md, Georgi:1992dw}: $m_{\pi} \geq 2\pi f_{\pi} /\sqrt{N_c}$.
Constraints on the ALP from NuCal (orange), CHARM (dark cyan), SLAC E-137 \& E-141 (gray), and SN1987A (light cyan at the bottom right corner) are also shown.
We convert the constraints to those on the $m_{\pi}$--$\alpha_{\pi}$ plane by assuming that $m_{\phi} = m_{\pi}$ and regarding $f$ as a function of $m_{\pi}$ and $f_{\pi}$ determined by the observed DM abundance.
The orange dotted line is the projected sensitivity of the Belle II experiment for an ALP search~\cite{Izaguirre:2016dfi}.
The projected SHiP experiment will examine the region below blue dashed lines, which covers large part of the parameter space where the observed DM abundance and the SIDM cross section are simultaneously achieved. 
}
\label{fig:constraints}
\end{figure*}

The coupling of Eq.~\eqref{eq:FFtilde}, which plays an important role in the kinetic equilibration between the pions and the SM plasma, is subject to various constraints on an ALP.
Among them, the most stringent constraints to our model come from beam dump experiments, which will be discussed shortly below.
We depict these constraints in the $m_{\pi}$--$\alpha_{\pi}$ plane of Fig.~\ref{fig:constraints}, by assuming that $m_{\phi} = m_{\pi}$ and $f$ is determined by the observed DM abundance as a function of $m_{\pi}$ and $f_{\pi}$. 
In the figure, we adopt $C_{\phi \gamma \gamma}=3$ as a representative value.

Beam dump experiments constrain an ALP by exploiting the production of ALPs through the Primakoff process of a virtual photon. 
Depending on the beam energy and the baseline distance, those experiments are sensitive to a particular range of the ALP lifetime and mass.
Proton beam dump experiments, such as CERN--Hamburg--Amsterdam--Rome--Moscow Collaboration(CHARM) ~\cite{Bergsma:1985qz} and NuCal~\cite{Blumlein:1990ay}, as well as electron beam dump experiments, such as Stanford Linear Accelerator Center (SLAC) E-137 \& E-141~\cite{Bjorken:1988as, Riordan:1987aw} constrain the coupling of a MeV-scale ALP to photons. 
Search for Hidden Particles (SHiP) is a projected proton beam dump experiment~\cite{Anelli:2015pba}.
(Projected) constraints of these experiments are plotted in Fig.~\ref{fig:constraints} based on Ref.~\cite{Dobrich:2015jyk}.
From Fig.~\ref{fig:constraints}, one can observe our model evades the existing constraints when $m_{\pi} = m_{\phi} \gtrsim 40 \MeV$. 
In particular, when the pion mass is around $100 \MeV$ and $\alpha_{\pi}$ is a few, our pion DM have a sizable self-scattering cross section that is compatible with SIDM hinted by the small scale crisis. 
The lower mass region ($m_{\pi} = m_{\phi} <100 \MeV$) is excluded mainly by NuCal, which has a relatively short baseline.
The lower bound on the mass tends to be relaxed as $C_{\phi \gamma \gamma}$ increases, since an produced ALP tends to decay before it reaches a {\it decay volume}.
Our model will be better probed by future beam dump experiments with increased sensitivity to a shorter-lived ALP.
Actually, as shown in Fig.~\ref{fig:constraints}, the SHiP experiment will be able to probe a substantial part of the parameter region where SIDM is realized in our model.
Furthermore, future $B$ factories such as the Belle II experiment~\cite{Abe:2010gxa} will cover a higher mass region ($m_{\pi} = m_{\phi} > 100 \MeV$)~\cite{Izaguirre:2016dfi} and be complementary to beam dump experiments as shown in Fig.~\ref{fig:constraints}.

Let us comment on an underlying model-dependent implication of an electroweak-scale ALP decay constant.
We can realize the ALP considered in this paper by introducing heavy vector-like fermions and scalar fields charged under the corresponding anomalous global symmetry as in the Kim--Shifman--Vainshtein--Zakharov axion model~\cite{Kim:1979if, Shifman:1979if}.
The heavy fermions carry a hypercharge and transform as some representation of SU$(N_{c})$ and the confining gauge group that are responsible for $V_{\rm UV}(\phi)$ in Eq.~\eqref{eq:L}.
In this case, the coupling of Eq. \eqref{eq:FFtilde} originates from an ALP coupling to the hyperchage gauge boson, and the Large Electron--Positron Collider (LEP) disfavors an ALP decay constant of $f/C_{\phi \gamma \gamma} \lesssim 3 \GeV$ (from the $Z \to \gamma \phi$, $\phi \to 2 \gamma$ decay) for an ALP with the mass around $100 \MeV$~\cite{Jaeckel:2015jla}, while the constraint may be improved by orders of magnitude in future lepton colliders such as the International Linear Collider (ILC)~\cite{Behnke:2013xla}, the Circular Electron Positron Collider (CEPC), and the Future Circular Collider (FCC-ee)~\cite{Gomez-Ceballos:2013zzn}.
In the case that the heavy fermions transform as some representation of the weak SU$(2)$ instead of the hypercharge, the induced flavor-changing neutral current process ($K^{\pm} \to \pi^{\pm} \phi$, $\phi \to 2 \gamma$) constrains the ALP decay constant: $f/C_{\phi \gamma \gamma} \gtrsim 170 \GeV$ for $m_{\pi} \sim 100 \MeV$~\cite{Izaguirre:2016dfi}.
Furthermore, the newly introduced heavy fermions may also be subject to collider constraints.
The heavy fermion mass is related with the ALP decay constant as $m_{\rm fermion} = Y f \sum_{r} 2 T(r)$ where $Y$ is a Yukawa coupling between the heavy vector-like fermions and the scalar fields.
We also define $T(r)$ by ${\rm Tr}(T_{r}^{i} T_{r}^{j}) = T(r) \delta_{ij}$ with the generator ($T_{r}$) of the representation ($r$), where we normalize the structure constant of SU$(N_{c})$ so that $T(r) = 1/2$ for the fundamental representation.
It follows that the heavy fermions may evade constraints from collider experiments, depending on a choice of the representation that may leads to $m_{\rm fermion}$ as heavy as $\sim 1 \TeV$ and/or the hypercharge.
Therefore, we just summarize the current status of searches for long-lived charged particles produced from the Drell-Yan process in the Large Hadron Collider (LHC): $m_{\rm fermion} > 650 \GeV$ with a charge unity in units of the positron charge~\cite{Khachatryan:2016sfv} and $m_{\rm fermion} > 310$ $(140) \GeV$ with a charge $2/3$ $(1/3)$~\cite{CMS:2012xi}.

Besides the collider experiments, an ALP with the mass around $100 \MeV$ is constrained also by supernovae (SN) since the energy loss rate of SN is enlarged by emission of ALPs~\cite{Masso:1995tw}. 
This constraint is relevant when the ALP mass is sufficiently light so that ALPs are thermally produced in a SN core where the temperature is about $50 \MeV$ and the coupling is in the range where ALPs are copiously produced but are not trapped in the core.
We find that in our model, the constraint from SN1987A becomes important only when $C_{\phi \gamma \gamma}$ is as small as $0.01$.

Finally, we would like to discuss aspects in indirect DM searches. 
In our model setup ($m_{\pi} \simeq m_{\phi}$), a pair of pions semi-annihilates into an ALP that subsequently decays into two photons with the energy of $m_{\pi} / 2$. 
Such a late-time semi-annihilation potentially affects, for example, Galactic and/or extra-Galactic gamma-rays (see, {\it e.g.}, Refs.~\cite{Jungman:1995df, Bergstrom:2000pn}) as well as the CMB anisotropies~\cite{Chen:2003gz, Padmanabhan:2005es}. 
Here remark that the semi-annihilation cross section is proportional to the relative velocity [see Eqs.~\eqref{eq:semisigma} and \eqref{eq:Idegegeratelim}] when $\Delta m = m_{\pi} - m_{\phi}$ is exactly zero.
Therefore, its effect at a later epoch of the Universe, when/where $v_{\rm rel} = v_{\rm rel, \, obs} < v_{\rm rel, \, fo}$ with $v_{\rm rel, \, fo}\simeq 2 \times 10^{5} \, {\rm km / s} \sqrt{19 / x_{\rm fo}}$ being the relative velocity at the pion freeze-out, is suppressed.

Let us take a closer look at the case that $m_{\pi} \gg \left| \Delta m \right| \neq 0$.
For $\Delta m > 0$, by comparing Eqs.~\eqref{eq:Imasslesslim} and \eqref{eq:Idegegeratelim}, we find that the semi-annihilation cross section scales as $\langle \sigma_{\rm semi} v_{\rm rel} \rangle = \langle \sigma_{\rm semi} v_{\rm rel} \rangle_{\rm fo} \, v_{\rm rel} / v_{\rm rel, \, fo}$ as as long as $v_{\rm rel}> v_{\rm rel, \, sat}=2\sqrt{|\Delta m|/m_{\pi}}$.
An observational upper bound on the semi-annihilation cross section denoted by $\langle \sigma_{\rm semi} v_{\rm rel} \rangle_{\rm obs}$ restricts our model to satisfy the following conditions: $v_{\rm rel, \, sat}$ and $v_{\rm rel, \, obs}$ should be smaller than $v_{\rm rel, \, fo} \left( \langle \sigma_{\rm semi} v_{\rm rel} \rangle_{\rm obs} / \langle \sigma_{\rm semi} v_{\rm rel} \rangle_{\rm fo} \right)$.
The first condition ($v_{\rm rel, \, sat} / v_{\rm rel, \, \rm fo} < \langle \sigma_{\rm semi} v_{\rm rel} \rangle_{\rm obs} / \langle \sigma_{\rm semi} v_{\rm rel} \rangle_{\rm fo}$) implies the required mass difference:
\bea
\frac{\Delta m}{m_{\pi}} < 0.07 \left( \frac{\langle \sigma_{\rm semi} v_{\rm rel} \rangle_{\rm obs}}{\langle \sigma_{\rm semi} v_{\rm rel} \rangle_{\rm fo}} \right)^{2} \left( \frac{19}{x_{\rm fo}} \right) .
\eea
The CMB anisotropies constrain the semi-annihilation cross section around and after the last scattering, where $v_{\rm rel, \, obs} \lesssim 2 \times 10^{-4} \, v_{\rm rel, \, fo} \sqrt{ 100 \MeV / m_{\pi} } \sqrt{ x_{\rm fo} / 19 }$, as $\langle \sigma_{\rm semi} v_{\rm rel} \rangle_{\rm obs} / \langle \sigma_{\rm semi} v_{\rm rel} \rangle_{\rm fo} \lesssim 0.01 \text{--} 0.1$ in the mass range of $m_{\pi} \simeq 0.1 \text{--} 1 \GeV$~\cite{Ade:2015xua, Slatyer:2015jla, Kawasaki:2015peu}.
The second condition ($v_{\rm rel, \, obs} / v_{\rm rel, \, fo} < \langle \sigma_{\rm semi} v_{\rm rel} \rangle_{\rm obs} / \langle \sigma_{\rm semi} v_{\rm rel} \rangle_{\rm fo}$) is trivially satisfied in this case, while the mass difference should be at maximum at a $10^{- (3 \text{--} 5)}$ level from the first condition.
A tighter bound is put by gamma-ray searches from the Galactic center (GC) in the Energetic Gamma Ray Experiment Telescope (EGRET)~\cite{Pullen:2006sy} and the Fermi Large Area Telescope (Fermi--LAT)~\cite{Ackermann:2015lka} for the DM mass larger than $100 \MeV$ (EGRET) and $200 \MeV$ (Fermi--LAT): $\langle \sigma_{\rm semi} v_{\rm rel} \rangle_{\rm obs} / \langle \sigma_{\rm semi} v_{\rm rel} \rangle_{\rm fo} \lesssim 10^{- (2 \text{--} 3)}$ for the isothermal profile.
Note that SIMP possesses a sizable self-scattering cross section and reduces the DM mass density in an inner part of halos~\cite{Yoshida:2000uw, Vogelsberger:2012ku, Rocha:2012jg, Vogelsberger:2015gpr}.
The first condition constrains the mass difference at a $10^{- (5 \text{--} 7)}$ level at maximum.
It is unclear whether the second condition is satisfied: $v_{\rm rel, \, obs} \lesssim 200 \text{--} 2000 \,  {\rm km / s} \, \sqrt{19 / x_{\rm fo}}$. 
This is because the DM velocity dispersion in the GC is poorly constrained (especially inside $10$\,kpc)~\cite{Battaglia:2005rj}.
Thus we do not show indirect detection constraints in Fig.~\ref{fig:constraints}.
Future cosmic gamma-ray searches with increased sensitivity to MeV-GeV photons such as e--ASTROGAM~\cite{DeAngelis:2016slk} may cover the lower mass region ($m_{\pi} < 100 \MeV$) where the observed DM abundance and the SIDM cross section are realized.

For $\Delta m  < 0$, the semi-annihilation is forbidden~\cite{DAgnolo:2015ujb} and the thermally averaged cross section is further suppressed effectively by a Boltzmann factor of $\exp \left( \Delta m / T \right) = \exp \left[ - \left( 4 / \pi \right) v_{\rm rel, \, sat}^{2} / v_{\rm rel}^{2} \right]$.
Let us assume that $\left| \Delta m \right| \ll T_{\rm fo} = m_{\pi} / x_{\rm fo}$, {\it i.e.}, $v_{\rm rel, \, sat} < 10^{5} \, {\rm km / s} \sqrt{19 / x_{\rm fo}}$, to ignore the suppression during the pion freeze-out and keep the discussion in the previous section intact.
In this case, the condition of 
\bea
\max \left[ v_{\rm rel, \, sat}, v_{\rm rel, \, obs} \right] \exp \left[ - \frac{4}{\pi} \frac{v_{\rm rel, \, sat}^{2}}{v_{\rm rel, \, obs}^{2}} \right]  \nonumber \\
< v_{\rm rel, \, fo} \left( \frac{\langle \sigma_{\rm semi} v_{\rm rel} \rangle_{\rm obs}}{\langle \sigma_{\rm semi} v_{\rm rel} \rangle_{\rm fo}} \right)
\eea
is not simplified unlike the case that $\Delta m  > 0$, though we can check if an observational constraint is satisfied in a case-by-case manner.
The left-hand side is a monotonically decreasing function of $v_{\rm rel, \, sat}$ and takes a maximum value of $v_{\rm rel, \, obs}$ at $v_{\rm rel, \, sat} = 0$.
As discussed above, when $\Delta m  = 0$, the observational constraints are satisfied so far, and thus no lower bound on $|\Delta m|$ is implied.
In other words, no fine tuning in the mass difference is required in this case.

\section{Conclusion}

In this paper, we presented a novel realization of SIMP as SIDM, where DM pions are associated with an ALP.  
Our model evaded shortcomings in the original SIMP model, such as an implicitly assumed mechanism of the kinetic equilibration between the pions and the SM thermal plasma and only marginal perturbativity. 
The former is solved by the ALP connecting the pions and the SM sector when the ALP mass is degenerate with the pion mass.
Meanwhile, the latter is alleviated because the chemical equilibration receives contribution from the semi-annihilation in addition to the $3\to2$ process and the semi-annihilation decouples later.

The newly introduced ALP is severely constrained by beam dump experiments. 
However, we have shown in a viable parameter region, DM pions possess a sizable self-scattering cross section, which (at least partially) solves the small scale crisis. 
Remarkably, most of the corresponding parameter region is within the reach of future beam dump experiments such as the SHiP experiment and will also be potentially probed by future cosmic gamma-ray searches such as e--ASTROGAM.

\begin{acknowledgements}
This work was supported by IBS under the project code, IBS-R018-D1.
We thank Sunghoon Jung, Kazunori Nakayama, Myeonghun Park, Brian Shuve, Chang Sub Shin, and Kazuya Yonekura for informative discussion.
\end{acknowledgements}

\clearpage
\appendix

\section{Generalization to other symmetry groups} \label{sec:app}

In the main text, we consider $N_{f}$ vector-like fermion pairs that transform as the fundamental and anti-fundamental representations of a SU$(N_{c})$ gauge group.
The model respects the global SU$(N_{f})_{L} \times$ SU$(N_{f})_{R}$ symmetry, except for a mass term that breaks its SU$(N_{f})_{A}$ subgroup explicitly.
When the chiral condensation forms, the global symmetry is broken into its SU$(N_{f})_{V}$ subgroup, which is assumed to be an exact symmetry ensuring the longevity of the resultant pions.
Our discussion does not change qualitatively for other gauge groups: (A) SO$(N_{c})$ ($N_{c} \ge 4$) and (B) USp$(N_{c})$ ($N_{c} \ge 4$).\footnote{
In our notation, USp$(2) \cong$ SU$(2)$ and thus $N_{c}$ should be even. 
The skew-symmetric matrix is denoted by $\Omega$.
}
We can introduce $N_{f}$ copies of Weyl fermions that transform as the fundamental representation of the gauge group.
Note that $N_{f}$ should be even for the USp$(N_{c})$ gauge group so that the gauge group evades the {\it global anomaly}~\cite{Witten:1982fp}.
The Lagrangian density is invariant under SU$(N_{f})$, while a mass term is introduced so that only the following subgroup is respected: (A) SO$(N_{f})$ and (B) USp$(N_{f})$.
In such a model, the confinement and chiral condensation are expected to occur, and thus the low-energy theory can be described by a non-linear sigma model where pions reside in (A) SU$(N_{f})$ / SO$(N_{f})$ and (B) SU$(N_{f})$ / USp$(N_{f})$~\cite{Witten:1983tx}.
The effective Lagrangian density is similar to Eq.~(\ref{eq:perchiralL}), while the broken SU$(N_{f})$ generators should satisfy (A) $T^{a} = (T^{a})^{T}$ and (B) $T^{a} \Omega = \Omega (T^{a})^{T}$.
It follows that the group factors in flavor-averaged cross sections [see Eqs.~\eqref{eq:cannsigma} and \eqref{eq:selfsigma}], which are defined as follows, are different from one model to another as summarized in Table~\ref{tab:groupfac}:
\bea
\label{eq:groupfac2}
d^{2} &=& \sum_{abc} d_{abc}^{2} , \\
\{C+R\}^{2} &=& \sum_{abcd} \{C+R\}_{abcd}^{2} , \nonumber \\
&=& \sum_{abcd} c_{(abcd)}^{2} + 16 r_{(ab)(cd)}^{2} / 9 , \\
\{C+R\}D^{2} &=& \sum_{abcd} \{C+R\}_{abcd} D^{2}_{abcd} , \\ 
D^{4} &=& \sum_{abcd} (D^{2}_{abcd})^{2} , \\
t^{2} &=& \sum_{abcde} c_{[abcde]}^{2} , \\
\{CD\}^{2} &=& \sum_{abcde} (\{CD\}_{abcde})^{2} ,
\eea
where we define 
\bea
d_{abc} &=& {\rm Tr} (T^{a} \{ T^{b}, T^{c}\} ) / 2 , \\
r_{abcd} &=& \sum_{e} f_{ace} f_{bde} , \\
\{C+R\}_{abcd} &=& c_{(abcd)} + 4 r_{(ab)(cd)} / 3 , \\
c_{abcd} &=& {\rm Tr}(T^{a} T^{b} T^{c} T^{d}) , \\
c_{abcde} &=& {\rm Tr}(T^{a} T^{b} T^{c} T^{d} T^{e}) , \\
D^{2}_{abcd} &=& \sum_{e} \left[ d_{abe} d_{cde} / 3 \right. \nonumber \\
&& \left. - (d_{ace} d_{bde} + d_{ade} d_{bce}) \right] , \\
\{CD\}_{abcde} &=& 4 c_{(abcde)} + \sum_{f} \left[ c_{(abcf)} d_{def} / 4 \right. \nonumber \\
&& - \left( c_{(abdf)} d_{cef} + c_{(acdf)} d_{bef} \right. \nonumber \\
&& \left. + c_{(bcdf)} d_{aef} + c_{(abef)} d_{cdf} \right. \nonumber \\
&& \left. + c_{(acef)} d_{bdf} + c_{(bcef)} d_{adf} \right) \nonumber \\
&& + 2 \left( d_{abf} c_{(cdef)}  + d_{acf} c_{(bdef)} \right. \nonumber \\
\label{eq:groupfac}
&& \left. \left. + d_{bcf} c_{(adef)} \right) / 3 \right] ,
\eea
with $f_{abc}$ being the structure coefficients of the broken generators.

Another caveat should be taken into account for the WZW term.
The prefactor of the WZW term [$2 N_{c}$ in Eq.~\eqref{eq:WZW}] is determined to reproduce the quantum anomaly of the fermion flavor-symmetry~\cite{Wess:1971yu, Witten:1983tw}, which we let $2 k$ denote.
There is a factor of two in it in the case of the SU$(N_{c})$ gauge group since both $N$ and $\bar{N}$ contribute to the quantum anomaly.
We find that it is $2k=N_{c}$ in the other gauge groups.
The WZW term can be written as a action on a five-dimensional ball, the boundary of which is the 4-dimensional Minkowski spacetime:
\bea
\Gamma_{\rm WZW} = 2\pi k \nu, \quad \nu = - i \frac{1}{480 \pi^{3}} \int {\rm Tr} \left( U^{-1} d U \right)^{5} \,,
\eea
with $d$ being the exterior derivative.
$U$ is parametrized by pions as in Eq.~(\ref{eq:piondef}) and given by (A) $V V^{T}$ and (B) $V \Omega V^{T} \Omega^{T}$ with $V$ being SU$(N_{f})$-valued fields.
For the WZW term to be independent of a choice of a 5-dimensional ball and thus well-defined, the above WZW action on a five-dimensional sphere should be a multiple of $2\pi$.
In fact, $\nu$ on a five-dimensional sphere measures the winding number of $U$ and thus takes a integer value.研究論文
Remark that $k$ takes a half-integer value ($N_{c}/2$) for the SO$(N_{c})$ gauge group, while it is integer ($N_{c}/2$, but $N_{c}$ is even) for the USp$(N_{c})$ gauge groups.
The WZW term for the SO$(N_{c})$ gauge group with an odd $N_{c}$ is, on the other hand, well-defined since the winding number of $U=V V^{T}$ is twice that of $V$.\footnote{
A similar discussion is found in Ref.~\cite{Auzzi:2008hu}.
} 
\clearpage

\begin{longtable*}{c|cc}
\hline \hline
G/H & $N_{\pi}$ & $d^{2}$ \\
\hline
SU$(N_{f})_{L} \times$ SU$(N_{f})_{R}$/SU$(N_{f})_{V}$  & $N_{f}^{2} - 1$ & $4 (N_{f}^{2} - 4) (N_{f}^{2} - 1) / N_{f}$ \\
SU$(N_{f})$/SO$(N_{f})$ & $(N_{f} - 1) (N_{f} + 2) / 2$ & $(N_{f} - 1) (N_{f}+4) (N_{f}^{2} - 4) / N_{f}$ \\
SU$(N_{f})$/USp$(N_{f})$ & $(N_{f} - 2) (N_{f} + 1) / 2$ & $(N_{f} - 4) (N_{f} + 1) (N_{f}^{2} - 4) / N_{f}$ \\
\hline \hline
\end{longtable*}

\begin{longtable*}{cc}
\hline \hline
$\{C + R\}^{2}$ & $\{C+R\}D^{2}$ \\
\hline
$8 (N_{f}^{2}-1) (3 N_{f}^{4} - 2 N_{f}^{2} + 6) / N_{f}^{2}$ & $16 (N_{f}^{2} - 4) (N_{f}^{2} - 1) (N_{f}^{2}+10) / (3 N_{f}^{2})$ \\
$(N_{f} - 1) (N_{f} + 2) (3 N_{f}^{4} + 7 N_{f}^{3} - 2 N_{f}^{2} - 12 N_{f} + 24) / N_{f}^{2}$ & $2 (N_{f} - 1) (N_{f} + 4) (N_{f}^{2} - 4) (N_{f}^{2} - 5 N_{f} + 20) / (3 N_{f}^{2})$ \\
$(N_{f} - 2) (N_{f} + 1) (3 N_{f}^{4} - 7 N_{f}^{3} - 2 N_{f}^{2} + 12 N_{f} + 24) / N_{f}^{2}$ & $2 (N_{f} - 4) (N_{f} + 1) (N_{f}^{2} - 4) (N_{f}^{2} + 5 N_{f} + 20) / (3 N_{f}^{2})$ \\
\hline \hline
\end{longtable*}

\begin{longtable*}{cc}
\hline \hline
$D^{4}$ & $t^{2}$ \\
\hline
$32 (11 N_{f}^{2} - 56) (N_{f}^{2} - 4) (N_{f}^{2} - 1) / (9 N_{f}^{2})$ & $4 (N_{f}^{2} - 4) (N_{f}^{2} - 1) N_{f} / 3$ \\
$4 (N_{f} - 1) (N_{f} + 2) (11 N_{f}^{2} + 25N_{f} -112) (N_{f}^{2} - 4) / (9 N_{f}^{2})$ & $(N_{f}^{2} - 4) (N_{f}^{2} - 1) N_{f} / 12$ \\
$4 (N_{f} - 2) (N_{f} + 1) (11 N_{f}^{2} - 25N_{f} - 112) (N_{f}^{2} - 4) / (9 N_{f}^{2})$ & $(N_{f}^{2} - 4) (N_{f}^{2} - 1) N_{f} / 12$ \\
\hline \hline
\end{longtable*}

\begin{longtable*}{c}
\hline \hline
$\{CD\}^{2}$ \\
\hline
$2 (N_{f}^{2} - 1) (N_{f}^{2} - 4) (833 N_{f}^{4} - 6630N_{f}^{2} + 11682) / (27 N_{f}^{3})$ \\
$(N_{f} - 1) (N_{f} + 4) (N_{f}^{2} - 4) (833 N_{f}^{4} + 3381 N_{f}^{3} - 10614 N_{f}^{2} - 20484 N_{f} + 46728) / (216 N_{f}^{3})$ \\
$(N_{f} - 4) (N_{f} + 1) (N_{f}^{2} - 4) (833 N_{f}^{4} - 3381 N_{f}^{3} - 10614 N_{f}^{2} + 20484 N_{f} + 46728) / (216 N_{f}^{3})$ \\
\hline \hline
\end{longtable*}

\begin{longtable*}{c}
\caption{\label{tab:groupfac}
Group factors in flavor-averaged cross sections of pions [Eqs.~(\ref{eq:semisigma})-(\ref{eq:cannsigma})] residing in different quotient spaces (G/H).
}
\end{longtable*}


\end{document}